\documentclass[%
reprint,
 amsmath,amssymb,
 aps,
]{revtex4-2}

\usepackage{graphicx}
\usepackage{dcolumn}
\usepackage{bm}

\usepackage{babel}

\usepackage{natbib}

\usepackage{letltxmacro}

\LetLtxMacro{\ORIGselectlanguage}{\selectlanguage}
\makeatletter
\DeclareRobustCommand{\selectlanguage}[1]{%
  \@ifundefined{alias@\string#1}
    {\ORIGselectlanguage{#1}}
    {\begingroup\edef\x{\endgroup
       \noexpand\ORIGselectlanguage{\@nameuse{alias@#1}}}\x}%
}
\newcommand{\definelanguagealias}[2]{%
  \@namedef{alias@#1}{#2}%
}
\makeatother

\definelanguagealias{en}{english}

\usepackage{amssymb}

\usepackage{float}

\usepackage{comment}

\begin{document}

\title{A quasiparticle-injection superconducting microwave relaxation oscillator}

\author{Giacomo Trupiano}
    \email[Corresponding author: ]{giacomo.trupiano@sns.it}
    \affiliation{NEST, Istituto Nanoscienze-CNR and Scuola Normale Superiore, Piazza S. Silvestro 3, I-56127 Pisa, Italy}

\author{Giorgio De Simoni}
     \email[Corresponding author: ]{giorgio.desimoni@nano.cnr.it}
    \affiliation{NEST, Istituto Nanoscienze-CNR and Scuola Normale Superiore, Piazza S. Silvestro 3, I-56127 Pisa, Italy}

\author{Francesco Giazotto}
    \email[Corresponding author: ]{francesco.giazotto@sns.it}
    \affiliation{NEST, Istituto Nanoscienze-CNR and Scuola Normale Superiore, Piazza S. Silvestro 3, I-56127 Pisa, Italy}
    

\begin{abstract}

We propose a superconducting microwave relaxation oscillator based on a nanowire shunted by a resistor and an inductor controlled by quasiparticle injection from a tunnel junction positioned on it: the QUISTRON. This device exhibits relaxation oscillator behavior with DC voltage-controlled frequency tuning and DC current bias. The device frequency is modulated via the tunnel junction, which induces localized heating by injecting quasiparticles. This heating mechanism modulates the nanowire switching current, enabling relaxation oscillations when it falls below the bias current.
We demonstrate the device operating principles and characterize its performance across various parameters, including different choices of shunt resistor, shunt inductance, and bath temperature ranging from 20 mK to 1 K. The device showed oscillation with a frequency range approximately between 1 GHz and 10 GHz and total energy dissipation per cycle of $\sim100$ zJ.
Our results suggest that this design offers a promising platform for compact, tunable superconducting oscillators in the microwave spectrum with potential applications in quantum information processing, microwave technology, and ultra-low-power electronics. The straightforward frequency control mechanism and integration potential make this device an attractive candidate for superconducting microwave local oscillators.
\end{abstract}

\keywords{superconducting, oscillator}

\maketitle

\section{Introduction} \label{sec:introduction}

\begin{figure*}[t!]
\includegraphics[width=\linewidth]{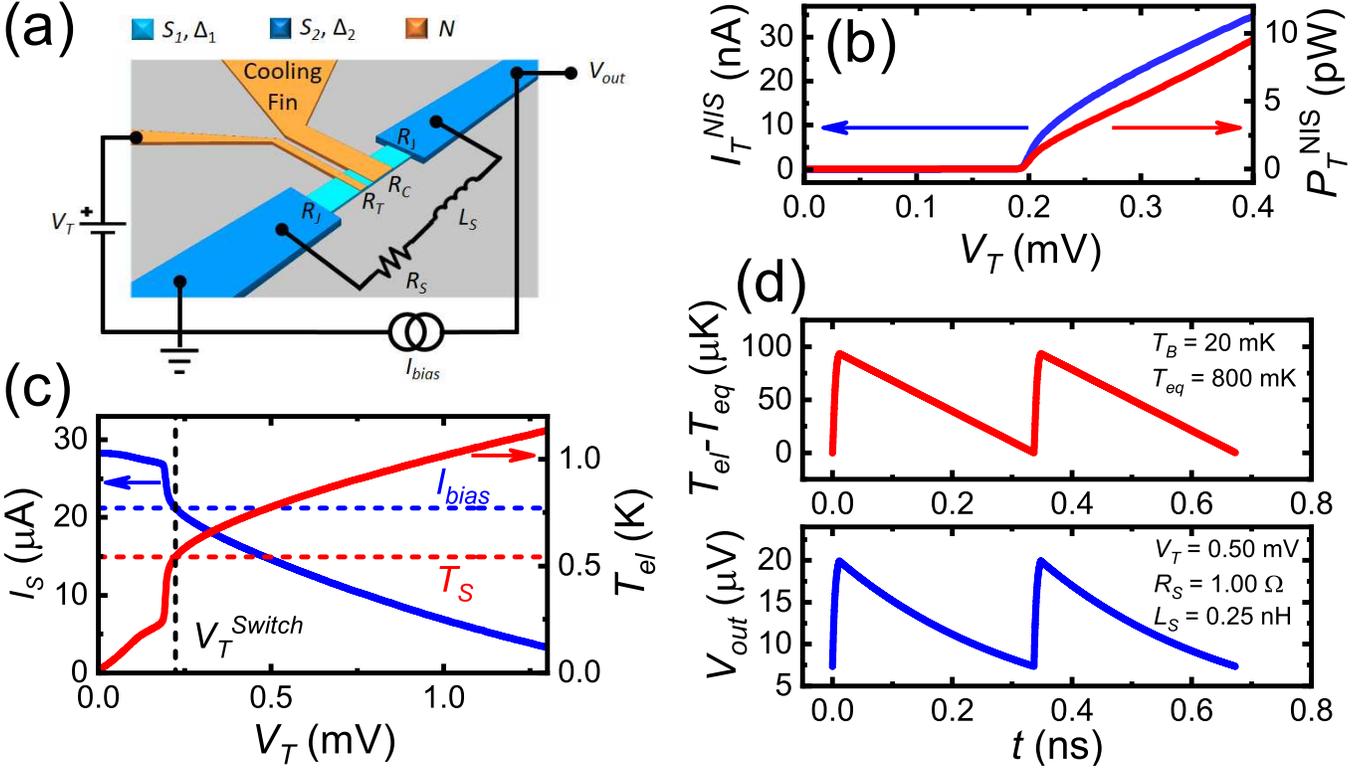}
\caption{\label{fig:fig1} QUISTRON scheme and operation. (a) Circuit scheme: A superconducting nanowire ($S_1$, gap $\Delta_1$, \textit{e.g.} aluminum) is shunted by a resistor $R_S$ and an inductor $L_S$, connected to superconducting banks ($S_2$, $\Delta_2 > \Delta_1$, \textit{e.g.} niobium) via tunnel junctions with resistance $R_J$. Additional tunnel junctions with resistances $R_T$ and $R_C$ are placed on the nanowire for quasiparticle injection and cooling, respectively, connected to a normal metal ($N$) contact and a cooling fin. The injection tunnel junction ($R_T$) heats the nanowire to enable relaxation oscillations. In contrast, the junction connected to the cooling fin ($R_C$) dissipates heat generated by Joule heating once the nanowire enters the normal state. (b) Normal metal-Insulator-Superconductor (NIS) junction characteristics: Charge ($I_T$, blue, left $y$-axis) and heat ($P_T$, red, right $y$-axis) currents vs. $V_T$ at bath temperature $T_B = 20$ mK. The NIS junction is insulating for subgap voltages; current increases rapidly at $V_T \simeq \Delta_1/e = 0.2$ mV. (c) Nanowire switching current ($I_S$, continuous blue line, left $y$-axis) and electronic temperature ($T_{el}$, continuous red line, right $y$-axis) vs. $V_T$. $I_S$ drops sharply at $V_T \simeq \Delta_1/e$ due to increased heat current and electronic temperature. The blue dashed line indicates the $I_{bias}$ value used in simulations shown in Fig. 4 and Fig. 5, and the black dashed line indicates switching voltage $V_T^{Switch}$, which is the voltage needed to have $I_S = I_{bias}$. The red dashed line indicates the switching electronic temperature $T_S$. (d) Simulated $T_{el}$ and voltage oscillations at $V_T = 0.50$ mV, $T_B = 20$ mK, $R_S = 1$ $\Omega$, $L_S = 0.25$ nH. $T_{el}$ spikes and cooling correspond to the nanowire switching between normal and superconducting states. Voltage oscillations result from the current redistribution between the nanowire and the shunt when the nanowire changes its state. The inverse of the oscillation period is $\nu_0 = 3$ GHz.
}
\end{figure*}
Superconducting electronics has emerged as a promising platform for quantum information processing \cite{RevModPhys.93.025005,doi:10.1126/science.1231930}, ultra-low-power computing \cite{likharev_rsfq_1991,Mukhanov_ERSFQ,chen_adiabatic_2019}, and high-sensitivity detectors \cite{irwin_application_1995,SNSPD_Goltsman,day_broadband_2003,giazotto2008ultrasensitive,guarcello2019josephson}. Within this field, superconducting oscillators play a crucial role in various applications, including local control of superconducting qubits \cite{krantz_quantum_2019, PhysRevLett.113.220502, 1211766, bao_cryogenic_2024} and clock generators for superconducting digital circuits \cite{403302, 717950}.

Traditionally, Josephson-junction-based oscillators have been the primary choice for superconducting oscillator applications \cite{koshelets_integrated_2000,galin_towards_2015}. These devices offer notable advantages, including low power consumption and compatibility with superconducting circuits. However, they often face significant challenges in terms of power output, necessitating large junction arrays. Although practical, this approach introduces scalability issues \cite{464849, galin_towards_2015}.

Recent advancements in superconducting nanowire technology have opened new avenues for device design. Superconducting nanowires have demonstrated remarkable properties, including high \cite{PhysRevApplied.11.044014, PhysRevApplied.20.044021, PhysRevApplied.13.054051, PhysRevLett.122.010504} and tunable \cite{annunziata_tunable_2010,vissers_frequency-tunable_2015} kinetic inductance, strong non-linearity \cite{PhysRevApplied.19.034024, PhysRevApplied.18.064088}, and sensitivity to local thermal perturbations \cite{allmaras_thin-film_2018, dane_self-heating_2022}. These characteristics make them attractive candidates for novel oscillator designs.

Using thermal effects to control superconducting devices has gained traction in recent years \cite{giazotto_opportunities_2006,fornieri2017towards}. Notably, the development of superconducting bolometers \cite{kokkoniemi_bolometer_2020, gunyho_single-shot_2024, paolucci_development_2020, paolucci_fully_2021}, hot-electron detectors \cite{shurakov_superconducting_2015}, superconducting nanowire single-photon detectors \cite{SNSPD_Goltsman, korzh_demonstration_2020,grunenfelder_fast_2023, charaev_single-photon_2024}, superconducting logic \cite{1062499,booth_quatratran_2000,mccaughan_superconducting-nanowire_2014, baghdadi_multilayered_2020,buzzi_nanocryotron_2023}, and neuromorphic \cite{toomey_design_2019, toomey_superconducting_2020, lombo_superconducting_2022} circuits based on controlled heating of superconducting nanowires have demonstrated the potential of thermally actuated superconducting devices. On this background, superconducting relaxation oscillators were demonstrated \cite{toomey_frequency_2018,toomey_microwave_2017}, exploiting a design incorporating a superconducting nanowire shunted by a resistor and biased above its switching current.

Here, we propose a superconducting microwave relaxation oscillator based on a nanowire shunted by a resistor and an inductor, which we call QUISTRON. Its operation leverages the thermal sensitivity of the superconducting nanowires and the controllable heat injection through a normal metal-insulator-superconductor (NIS) tunnel junction \cite{giazotto_josephson_2005, lemziakov_applications_2024,tirelli2008manipulation} overlapped with the nanowire.
Our approach allows for a compact design without needing large arrays of Josephson junctions \cite{464849,galin_towards_2015} and, differently from previous nanowire-based oscillators, it takes advantage of the quasiparticle injection and of the localized heating to modulate the nanowire switching current. A voltage-controlled quasiparticle injection scheme allows us to achieve a large-bandwidth tuning range without complex flux-based control schemes. 

Furthermore, the simple DC control and the compatibility with standard superconducting fabrication processes may facilitate an easy integration with other superconducting circuit elements. Finally, by leveraging the ultra-low dissipation of superconducting nanowires, the QUISTRON promises an energy efficiency comparable to or better than state-of-the-art superconducting oscillators \cite{koshelets_integrated_2000,galin_towards_2015} and to address several challenges in the field of in quantum information processing and of superconducting electronics. 
Indeed, it could serve as a local oscillator for superconducting qubits \cite{krantz_quantum_2019, PhysRevLett.113.220502, 1211766, bao_cryogenic_2024}, or as a clock generator for superconducting digital circuits \cite{403302, 717950}. Its tunability could be advantageous in superconducting parametric amplifiers \cite{PhysRevApplied.21.014052}.
Moreover, the device capability to generate multiple harmonics envisages its employment as a frequency source for multiplexed readout systems. This feature is precious for transition edge sensors (TES) \cite{irwin_application_1995, lanting_frequency-domain_2005} and kinetic inductance devices (KID) \cite{day_broadband_2003,giazotto2008ultrasensitive,sipola_multiplexed_2019}, both of which play crucial roles in many astrophysics applications \cite{paolucci_development_2020, paolucci_fully_2021,mauskopf_transition_2018}.

In the following sections, we present the device concept, the theoretical framework, and the results of numerical simulations. We investigate device performance across various parameters, including shunt resistance, inductance values, and operating temperature. Finally, we discuss the potential applications and prospects of the QUISTRON in the context of quantum information processing, microwave technology, and ultra-low-power electronics.
 
\section{Operation of the device} \label{sec:device}

\begin{figure*}[t!]
\includegraphics[width=\linewidth]{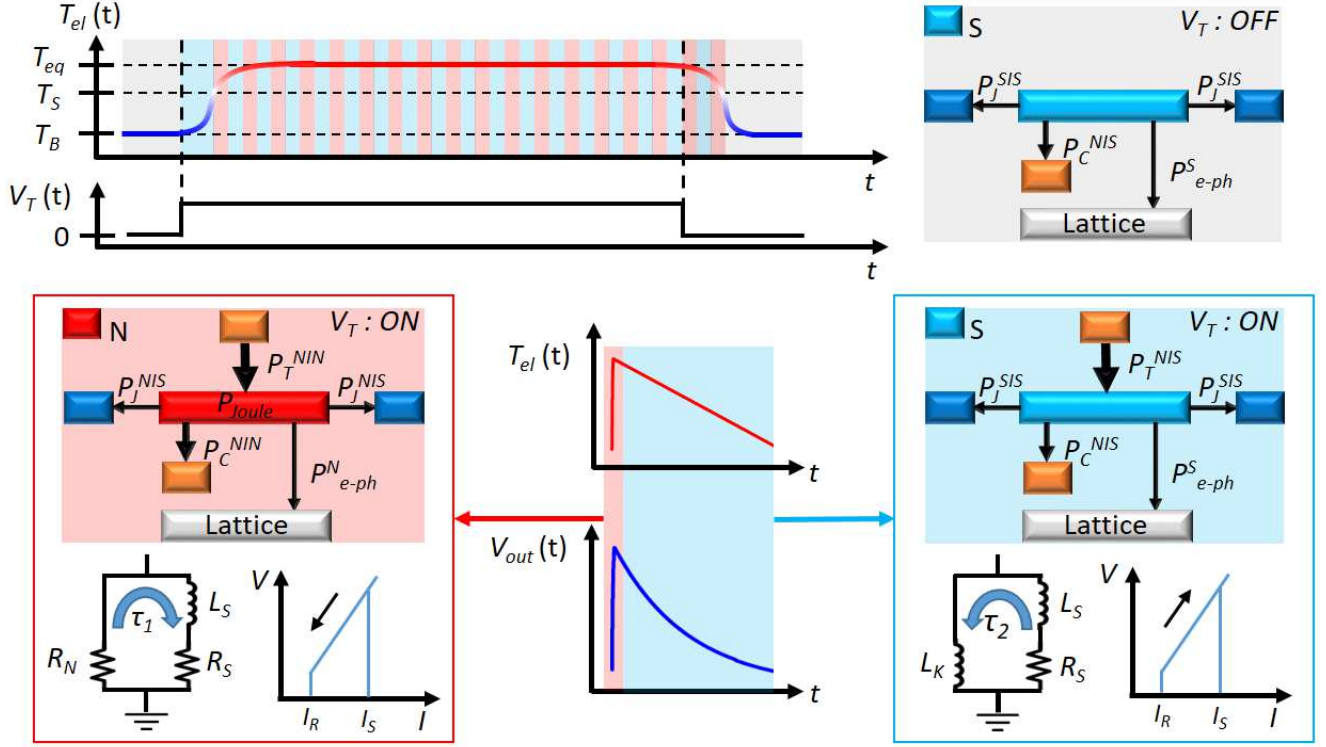}
\caption{\label{fig:fig2} Schematic of the QUISTRON operating principle, heat currents, and voltage and temperature oscillations. The color code indicates nanowire states and heat currents: blue (superconducting, voltage ON), red (normal, voltage ON), and grey (superconducting, voltage OFF). Oscillations are controlled by a tunnel junction voltage ($V_T$) applied to the nanowire. When $V_T = 0$, the nanowire electronic temperature ($T_{el}$) is at equilibrium with the bath temperature ($T_B$), maintaining superconductivity. As $V_T$ exceeds $\Delta_1/e$ ($\Delta_1$: superconducting gap), $T_{el}$ increases due to quasiparticle injection (blue to red color gradient in the upper-left scheme). When $V_T$ surpasses a critical value $V_T^{Switch}$, $T_{el}$ exceeds the switching temperature $T_S$, transitioning the nanowire to its normal state. The bias current redirects to the shunt resistor $R_S$ with characteristic time $\tau_1 = L_S/(R_S + R_N)$, where $L_S$ is the shunt inductance, $R_S$ the shunt resistance, and $R_N=R_{NW}+2 R_J$ is the total normal state resistance of the series of the two Josephson junctions $R_J$ and the nanowire. This generates $V_{out}$ and further heats the nanowire. As the nanowire current falls below the retrapping current $I_R$, it reverts to the superconducting state. Current then redirects back to the nanowire with characteristic time $\tau_2 = (L_S + L_K)/R_S$, where $L_K$ is the kinetic inductance. Consequently, $V_{out}$ decreases. $T_{el}$ rises during oscillations until reaching equilibrium temperature $T_{eq}$, where heating during the normal state balances cooling while in the superconducting state. This process repeats cyclically, producing periodic oscillations in $V_{out}$ and $T_{el}$ until $V_T$ is set to zero. The nanowire oscillates while $T_{el}$ gradually decreases. When $T_{el}$ falls below $T_S$, oscillations cease, and $T_{el}$ returns to the bath temperature $T_B$ (red to blue color gradient in the upper-left scheme).
}
\end{figure*}
Fig.~\ref{fig:fig1}(a) shows the core of the QUISTRON: a superconducting nanowire ($S_1$) with a superconducting gap $\Delta_1$, shunted by a resistor $R_S$ and an inductor $L_S$. The nanowire connects to superconducting banks ($S_2$) with a larger superconducting gap ($\Delta_2 > \Delta_1$) via tunnel junctions (resistance $R_J$). The key in our design is the addition of two specialized tunnel junctions on the nanowire. The first, with resistance $R_T$, connects to a normal metal ($N$) electrode and is the quasiparticle injection point for heating the nanowire. The second, with resistance $R_C$, connects to a cooling fin to dissipate heat generated when the nanowire transitions to the normal state.
In this work, we assume the nanowire to be made of aluminum, with a gap $\Delta_1 = 0.2$ meV and a critical temperature $T_{c1}=1.3$ K, and the banks to be made of niobium, with a gap $\Delta_2=1.4$ meV and a critical temperature $T_{c2}=9.2$ K.
The operation of the device relies on the controlled injection of quasiparticles through the tunnel junction to regulate the electronic temperature of the nanowire. Fig.~\ref{fig:fig1}(b) illustrates the NIS (Normal metal-Insulator-Superconductor) junction characteristics, showing charge ($I_T^{NIS}$) and heat ($P_T^{NIS}$) currents versus applied voltage $V_T$ at a bath temperature of 20 mK. The NIS junction is insulating for subgap voltages, with currents increasing steeply at $V_T \simeq \Delta_1/e = 0.2$ mV, where $e$ is the elementary charge. The $S_2$ banks play a crucial role in enhancing the device efficiency by confining the injected quasiparticles within the nanowire. This confinement occurs because $\Delta_2 > \Delta_1$, causing the $S_2IS_1$ junctions to act as Andreev mirrors for the quasiparticles \cite{giazotto_opportunities_2006,sandgren_fabrication_2002}.
Fig.~\ref{fig:fig1}(c) shows how the nanowire switching current ($I_S$) and electronic temperature ($T_{el}$) vary with $V_T$. $I_S$ drops sharply at $V_T \simeq \Delta_1/e$ due to increased heat current and $T_{el}$. This graph also indicates the bias current ($I_{bias}$) used in our simulations and the switching voltage ($V_T^{Switch}$), which is the voltage needed to have $I_S = I_{bias}$.
Fig.~\ref{fig:fig1}(d) shows the voltage and temperature oscillations, which are controlled by the tunnel junction voltage ($V_T$). When $V_T$ exceeds a critical value, the nanowire electronic temperature ($T_{el}$) rises above the switching temperature ($T_S$), causing a transition to the normal state. This redirects the bias current to the shunt resistor, generating $V_{out}$ and further heating the nanowire. As the nanowire current drops below the retrapping current $I_R$, it returns to the superconducting state, redirecting current back to the nanowire. This cycle repeats, creating periodic oscillations in $V_{out}$ and $T_{el}$ until $V_T$ is set to zero. The process involves characteristic times $\tau_1$ and $\tau_2$, which depend on various circuit parameters. This mechanism is explained in detail in the next paragraph and illustrated in Fig.~\ref{fig:fig2}.

The oscillation microscopic mechanism is illustrated in Fig.~\ref{fig:fig2}. When $V_T = 0$, the nanowire electronic temperature ($T_{el}$) is at equilibrium with the bath temperature ($T_B$), maintaining superconductivity (Fig.~\ref{fig:fig2}, grey regions). As $V_T$ exceeds $\Delta_1/e$, $T_{el}$ increases due to quasiparticle injection (Fig.~\ref{fig:fig2}, blue regions). When $V_T$ surpasses the critical value $V_T^{Switch}$, $T_{el}$ exceeds the switching temperature $T_S$, transitioning the nanowire to its normal state. This causes the bias current to redirect toward the shunt resistor $R_S$ with a characteristic time $\tau_1 = L_S/(R_S + R_N)$, where $L_S$ is the shunt inductance and $R_N = R_{NW}+2 R_J$ is the total normal-state resistance of the series of the two Josephson junctions $R_J$ and the nanowire. This generates the output voltage $V_{out}$ and further heats the nanowire (Fig.~\ref{fig:fig2}, red regions). As the nanowire current falls below the retrapping current $I_R$, it reverts to the superconducting state. The current then redirects back to the nanowire with characteristic time $\tau_2 = (L_S + L_K)/R_S$, where $L_K$ is the total kinetic inductance of the series of the two Josephson junctions $R_J$ and the nanowire, causing $V_{out}$ to decrease (Fig.~\ref{fig:fig2}, blue regions).
This process repeats cyclically, producing periodic oscillations in $V_{out}$ and $T_{el}$, as shown in Fig.~\ref{fig:fig1}(d). The simulation shown in Fig.~\ref{fig:fig1}(d) was performed at $V_T=0.50$ mV, $T_B = 20$ mK, $R_S=1$ $\Omega$, and $L_S=0.25$ nH, and it shows $T_{el}$ spikes corresponding to the nanowire switching between normal and superconducting states. The resulting voltage oscillations have a $\nu_0 = 3$ GHz frequency. The temperature and voltage oscillations continue until $V_T$ is set to zero (Fig.~\ref{fig:fig2}, grey region), causing $T_{el}$ to decrease gradually. When $T_{el}$ falls below $T_S$, oscillations cease, and $T_{el}$ returns to $T_B$.

The heat transport mechanisms are represented in schemes within colored regions. The blue area depicts the heat currents scheme when the nanowire is superconducting and the tunnel junction is injecting, where $P_T^{NIS}(T_{el}, V_T)$ is the heat current of the injection junction, $P_C^{NIS}(T_{el})$ is the heat current flowing through the cooling fin, $P_J^{SIS}(T_{el})$ is the heat current through the leads, and $P_{e-ph}^S(T_{el})$ is the electron-phonon coupling heat current.
The red region shows the heat currents scheme when the nanowire is in the normal state and the tunnel junction is injecting, where $P_T^{NIN}(T_{el}, V_T)$ is the heat current of the injection junction, $P_{Joule}$ is the Joule heating power, $P_C^{NIN}(T_{el})$ is the heat current flowing through the cooling fin, $P_J^{SINIS}(T_{el}, V_{out})$ is the heat current through the series of the two junctions and the nanowire, and $P_{e-ph}^N(T_{el})$ is the electron-phonon coupling heat current.
The grey region illustrates the heat currents scheme when the nanowire is superconducting, and the tunnel junction is not injecting. These heat currents were used to simulate the thermal dynamics of the device, as described in the following section. The detailed calculations of these heat currents are provided in Appendix \ref{sec:appendix1}.

The characteristic times $\tau_1$ and $\tau_2$ represent the timescales for current redistribution between the nanowire and shunt resistor, influencing the shape and frequency of the oscillations. The oscillation frequency depends on the characteristic $L/R$ time of the circuit and $I_{bias}$, $I_S$, and $I_R$. An approximated relation for the period of a single relaxation oscillation is \cite{toomey_frequency_2018,toomey_microwave_2017}:
\begin{align}
T = \frac{1}{\nu_0}\simeq\tau_1 \log{\left(\frac{I_S}{I_R}\right)} + \tau_2 \log{\left(\frac{I_{bias}-I_R}{I_{bias}-I_S}\right)},
\label{eq:Period}
\end{align}
where all quantities dependent on the electronic temperature $T_{el}$, such as $\tau_2$ (through $L_K$) and $I_S$, are evaluated at the equilibrium temperature $T_{eq}$. For this reason, it is possible to modulate the frequency by changing the voltage applied to the injection junction, $V_T$. By increasing $V_T$, we can increase the power injected and, thus, the electronic temperature of the nanowire. Since $I_S$ monotonically decreases with increasing $T_{el}$, while $I_{bias}$ remains constant, the oscillation frequency increases accordingly. 

A similar mechanism was previously exploited to realize a superconducting relaxation oscillator based on a shunted superconducting nanowire \cite{toomey_frequency_2018,toomey_microwave_2017} current biased above its switching current. In this implementation, the oscillation frequency was modulated by increasing or decreasing the bias current with respect to the switching current. The main difference in our approach is that we can control the oscillation by applying a voltage to the injection junction while keeping $I_{bias}$ constant. This allows us to control the relaxation oscillation using a much smaller current. Indeed, as shown in Fig.~\ref{fig:fig1}(c), it is sufficient to apply $V_T \simeq 0.25$ mV, which corresponds to the injection of a tunnel current $I_T \lesssim 20$ nA, as shown in Fig.~\ref{fig:fig1}(b). If we used the bias current to control the oscillation frequency, we would have to change it by a sizable portion of the switching current, around 10 $\mu$A.

\section{Circuit Model and Relaxation Oscillations Simulation}\label{sec:methods}

\subsection{Circuit Design and Parameter Choices}

The circuit parameters used in the calculations were: $R_T = 10$ k$\Omega$, $R_C = 500$ $\Omega$, and $R_J = 10$ $\Omega$, while we simulated the device for different choices of $R_S$ and $L_S$, as shown in Fig.~\ref{fig:fig4} and Fig.~\ref{fig:fig5}. The choices of $R_T$ and $R_C$ allowed to achieve good electronic input-output insulation between the nanowire and the injection junction ($R_T \gg R_N$), efficient heating while the nanowire is superconducting ($P_T^{NIS} \gg P_C^{NIS}$), and effective heat dispersion through the cooling fin when the nanowire is in the normal state ($P_C^{NIN} \sim P_T^{NIN}$, $P_{Joule}$).

The cooling fin provides an innovative and simple method for introducing passive feedback control on the electronic temperature of the nanowire. This is because the heat current flowing through the junction connected to the cooling fin is heavily dependent on the state of the nanowire. Given that the gapped density of states in the superconductor suppresses quasiparticle tunneling, we have $P_C^{NIS} \ll P_C^{NIN}$. Consequently, the heat current dispersed through the cooling fin is negligible when the nanowire is superconducting, enhancing the heating process efficiency. Conversely, when the nanowire is in the normal state, the heat current dispersed through the cooling fin is comparable to the Joule heating, thus preventing runaway heating of the nanowire.

The nanowire switching current dependence was estimated using the Bardeen equation for a dirty superconductor \cite{RevModPhys.34.667}:
\begin{align}
I_S(T_{el}) = \frac{J_c(0) w t}{2\sqrt{2}}\left[1-\left(\frac{T_{el}}{T_{c1}}\right)^{2}\right]^{3/2},
\label{eq:Bardeen}
\end{align}
where $J_c(0) = 100$ GA/m$^2$ is the nanowire critical current density at 0 K, $w=40$ nm is its width, and $t=20$ nm is its thickness. The critical current density value is extracted by \cite{PhysRevB.26.3648}, where the $J_c$ of aluminum films of different thicknesses is analyzed down to $t=20$ nm. With this choice of parameters, the maximum switching current of the aluminum nanowire is estimated to be $I_S^{max}\simeq 28$ $\mu$A. The film resistivity is also provided in the same work \cite{PhysRevB.26.3648}. In particular, the resistivity of a 20-nm-thick film is $\rho_{Al}\simeq 26$ n$\Omega$m. Using this result, we can extrapolate the normal-state resistance of the nanowire to be $R_{NW}=\rho_{Al}(l/wt)\simeq40$ $\Omega$, where $l=1.2$ $\mu$m is the length of the nanowire used in our calculation. Thus, the total normal-state resistance of the series combination of the two Josephson junctions $R_J$ and the nanowire is $R_N = R_{NW} + 2 R_J \simeq 60$ $\Omega$. The $R_J$ resistance was chosen to be equal to 10 $\Omega$ so that the critical current of the Josephson junction formed by the $S_2IS_1$ junctions ($I_c$) would be larger than the switching current of the nanowire. This requirement simplifies the dynamics, as we can treat the two Josephson junctions as (Andreev) mirrors for quasiparticles.

The Josephson junction critical current ($I_c$) and its temperature dependence were calculated using the Ambegaokar-Baratoff relation \cite{PhysRevLett.10.486,tinkham2004introduction}:
\begin{align}
I_c(T_{el}) = \frac{\pi \Delta_1(T_{el})}{2 e R_J} \tanh{\bigg[\frac{\Delta_1(T_{el})}{2k_B T_{el}}\bigg]},
\end{align}
where $k_B$ is the Boltzmann constant, and the gap dependence on the electronic temperature was estimated using the numerical approximation of the gap equation solution \cite{gross_anomalous_1986}:
\begin{align}
\Delta_1(T_{el}) \simeq \Delta_1 \tanh{\left(1.74\sqrt{\frac{T_{c1}}{T_{el}} - 1}\right)}.
\end{align}
With this choice of parameters, the maximum critical current of the Josephson junctions was estimated to be $I_c^{max}\simeq 31$ $\mu$A. It is important to note that the condition $I_c^{max}>I_S^{max}$ is not strictly necessary to ensure the correct operation of the device. This is because the nanowire switching current dependence on the electronic temperature is steeper than that of the Josephson critical current. It is sufficient that the bias current intersects the $I_S(T_{el})$ curve before the $I_c(T_{el})$ curve. This condition ensures that the nanowire switches to the normal state before the Josephson junctions lose their superconductivity so that the dynamics of the device is controlled only by the state of the nanowire. This relaxed condition provides more flexibility in the design and operation of the device, allowing for a broader range of parameters while still maintaining the core functionality of the QUISTRON.

The Josephson junctions were designed to be non-hysteretic to ensure their critical current and switching current were larger than the nanowire switching current. We used high-transparency Josephson junctions to achieve this, as shown in \cite{10.1063/1.2357915}. We assumed typical values for the specific capacitance $c_s = 50$ fF/$\mu$m$^2$ and a low but experimentally feasible specific resistance $r_s = 20$ $\Omega$/$\mu$m$^2$ \cite{10.1063/1.2357915}. With these parameters, we estimated a Stewart-McCumber parameter $\beta_c \simeq 0.9$, indicating the non-hysteretic behavior of the Josephson junction \cite{tinkham2004introduction}.

One possible design for the QUISTRON involves directly connecting the aluminum superconducting nanowire to the niobium banks without an insulating interface. Without the insulating interface, the larger superconducting gap of the banks would still serve as Andreev mirrors for the quasiparticles in the nanowire, effectively trapping the heat in the nanowire. However, we opted to include insulating interfaces with resistance $R_J$ in the device design. This approach simplified the modeling of the heat current between the nanowire and the banks by providing a well-defined interface and ensuring better thermal insulation. Additionally, it avoids the complications introduced by the proximity effect between the two superconductors \cite{usadel}. Nonetheless, this alternative design could be easier to implement and should be considered in future research.

\subsection{Steady-state simulations}\label{sec:steady}

Steady-state simulations were performed to understand the behavior of QUISTRON. These simulations aimed to identify two main parameters: the steady-state electronic temperature achieved by the superconducting nanowire when starting from its equilibrium state with the bath temperature, and the switching current based on the applied injection voltage.
The initial simulations focused on characterizing the heat current of the injection junction in relation to the injection voltage while the nanowire remained superconducting. Heat transfer calculations supported these simulations through three pathways: the leads, the cooling fin, and the electron-phonon coupling mechanism.
The overall heat balance was determined by adding the contributions from these pathways with their respective signs: heat current flowing into the nanowire was considered positive, while heat current flowing out of the nanowire was considered negative. 
To find the steady-state electronic temperature of the nanowire, $T_{el}$, as a function of $V_T$, we solved the heat balance equation for $T_{el}$:
\begin{align}
    &P_T^{NIS}(V_T, T_{el}) - P_J^{SIS}(T_{el}) \nonumber \\
    &\quad - P_C^{NIS}(T_{el}) - P_{e-ph}^{S}(T_{el}) = 0,
\end{align}
where $P_T^{NIS}(V_T, T_{el})$ is the heat current of the injection junction when the nanowire is superconducting, shown in Fig.~\ref{fig:fig1}(b), $P_J^{SIS}(T_{el})$ is the heat current flowing through the leads, $P_C^{NIS}(T_{el})$ is the heat current flowing through the cooling fin, and $P_{e-ph}^{S}(T_{el})$ is the heat current due to the electron-phonon coupling (see Appendix \ref{sec:appendix1} for details).
By solving this equation, we obtained the steady-state electronic temperature as a function of the applied injection voltage, $T_{el}(V_T)$, which is represented by the red curve in Fig.~\ref{fig:fig1}(c). Using this relationship and the Bardeen relation between the switching current and the electronic temperature, $I_S(T_{el})$, as shown in Eq.~(\ref{eq:Bardeen}), we can estimate the dependence of the switching current on the applied voltage, $I_S(V_T)$, represented by the blue curve in Fig.~\ref{fig:fig1}(c).

Once the $I_S(V_T)$ relation is known, we can estimate the voltage needed to switch the nanowire to the normal state for a given bias current $I_{bias}$. This is the voltage $V_T^{Switch}$ needed to lower $I_S$ to the value of $I_{bias}$. In this simulation, we chose a bias current that is $75\%$ of the switching current evaluated at the bath temperature ($T_B = 20$ mK). For this value of $I_{bias}$, we estimate $V_T^{Switch} = 0.22$ mV, which corresponds to a switching temperature $T_S = 0.54$ K, as it is shown in Fig.~\ref{fig:fig1}(c).

\subsection{Dynamics simulation}\label{sec:dynamics}

The relaxation oscillation dynamics was simulated by solving the heat equation of the nanowire. This took into account various heat currents and electronic heat capacity, both when the nanowire is superconducting and in the normal state (see Fig.~\ref{fig:fig2} for red regions and heat currents scheme). The simulation starts when the nanowire reaches the switching temperature $T_S$, causing it to transition to the normal state. 
In this state, the dynamics of the electronic temperature of the nanowire is described by the differential equation
\begin{align}
C_N \frac{dT_{el}}{dt} &= P_T^{NIN}(T_{el}, V_T) + P_{Joule} - P_C^{NIN}(T_{el}) \nonumber \\
&\quad - P_J^{SINIS}(T_{el}, V_{out}) - P_{e-ph}^N(T_{el}),
\label{eq:Dynamics_N}
\end{align}
with the initial condition for the temperature being $T_{el}(0) = T_S$, where $C_N$ is the nanowire electronic heat capacity in the normal state (see Appendix \ref{sec:appendix2} for details on its estimation), $P_T^{NIN}(T_{el}, V_T)$ is the heat current of the injection junction, $P_{Joule}$ is the Joule heating power, $P_C^{NIN}(T_{el})$ is the heat current flowing through the cooling fin, $P_J^{SINIS}(T_{el}, V_{out})$ is the heat current through the series of the two junctions and the nanowire, and $P_{e-ph}^N(T_{el})$ is the electron-phonon coupling heat current (see Appendix \ref{sec:appendix1} for details on their estimations and red shaded areas in Fig.~\ref{fig:fig2} for schematic representation of the heat currents).
In this equation, we neglect the spatial gradient of the temperature because, at the operation temperature ($T<T_{c1}$), the nanowire dimensions are smaller than the electron-phonon scattering length (see Appendix \ref{sec:appendix2} for details on its estimation). This condition ensures that the dynamics of the heat currents are approximately homogeneous along the entire nanowire.

To consistently calculate $P_J^{SINIS}(T_{el}, V_{out})$, and especially the heat due to the Joule effect $P_{Joule} = V_{out}(t)I_{NW}(t)$, we have to estimate the time dependence of $V_{out}(t)$ and $I_{NW}(t)$, which are the voltage applied across the nanowire and the charge current flowing through it. We know that the initial condition for the current flowing through the nanowire is $I_{NW}(0) = I_S(T(0) = T_S)$ because the nanowire switches to the normal state at $t = 0$. We assume that the transition to the normal state is instantaneous ($h/\Delta_1\sim 10$ ps, much smaller than other characteristic times $\tau_1$ and $\tau_2$) so that the circuit made by the shunted nanowire instantaneously becomes an RL circuit with time constant $\tau_1 = L_S/(R_N + R_S)$.  Thus, the time dependence of the current flowing through the nanowire and of the voltage drop across it are, respectively:
\begin{align}
I_{NW}(t) &= (I_S(T(0)) - I_{min})e^{-\frac{t}{\tau_1}} + I_{min}, \\
V_{out}(t) &= R_{\parallel}(I_{bias} - I_{NW}(t)),
\end{align}
where $I_{min} = I_{bias}R_{\parallel}/R_N$ is the minimum current that can flow through the nanowire when the circuit is at steady state, $R_{\parallel}=(1/R_{N}+1/R_S)^{-1}$ is the parallel resistance of the shunted nanowire, and $\tau_1$ is the RL characteristic time of the circuit.

While the nanowire is normal, its electronic temperature rises, primarily due to heating from the Joule effect. The simulation ends when the current flowing through the nanowire equals its retrapping current ($I_R$): $I_{NW}(t_{final}) = I_R$ (see Appendix \ref{sec:appendix3} for details on the estimation of the retrapping current). When this occurs, the nanowire instantaneously transitions to the superconducting state, thereby changing the dynamics of the heat and charge currents (Fig.~\ref{fig:fig2} blue regions and heat currents scheme). In this state, the dynamics of the electronic temperature of the nanowire is described by the differential equation:
\begin{align}
C_S \frac{dT_{el}}{dt} &= P_T^{NIS}(T_{el}, V_T) - P_C^{NIS}(T_{el}) \nonumber \\
&\quad - 2P_J^{SIS}(T_{el}) - P_{e-ph}^S(T_{el}),
\label{eq:Dynamics_S}
\end{align}
with the initial condition for the temperature being $T_{el}(0) = T_{final}$, where $T_{final}$ is the temperature reached at the end of the simulation described by Eq.~(\ref{eq:Dynamics_N}). Here, $C_S$ is the nanowire electronic heat capacity in the superconducting state (see Appendix \ref{sec:appendix2} for details on its estimation), $P_T^{NIS}(T_{el}, V_T)$ is the heat current of the injection junction, $P_C^{NIS}(T_{el})$ is the heat current flowing through the cooling fin, $P_J^{SIS}(T_{el})$ is the heat current through the leads, and $P_{e-ph}^S(T_{el})$ is the electron-phonon coupling heat current (see Appendix \ref{sec:appendix1} for details on their estimations and blue shaded areas in Fig.~\ref{fig:fig2} for schematic representation of the heat currents).

Also, in this case, the simulation of the thermal dynamics is bound to the dynamics of the current flowing through the nanowire. That is because the simulation described in Eq.~(\ref{eq:Dynamics_S}) ends when $I_{NW}=I_S$, causing the switching of the nanowire to the normal state. We know that the initial condition for the current flowing through the nanowire is $I_{NW}(0) = I_R$ because the nanowire switches to the superconducting state at $t = 0$. We assume, again, that the transition is instantaneous so that the circuit instantaneously becomes an RL circuit with time constant $\tau_2(T_{el}) = [L_S+L_K(T_{el})]/R_S$, where $L_K$ is the kinetic inductance of the series of the two junctions and the nanowire (see Appendix \ref{sec:appendix3} for details on its estimation). Thus, the time dependence of the current flowing through the nanowire and the voltage drop across it are, respectively:
\begin{align}
I_{NW}(t) &= (I_{bias} - I_{R})\left(1-e^{-\frac{t}{\tau_2(T_{el}(t))}}\right) + I_{R}, \\
V_{out}(t) &= R_{S}(I_{bias} - I_{NW}(t)),
\end{align}
where we approximated $\tau_2(T_{el}(t)) \simeq \tau_2(T_{el}(0))$, since the thermal oscillations ($\sim 100$ $\mu$K) are much smaller than the absolute temperature of operation ($\sim 100$ mK).

While the nanowire is superconducting, there are two possible dynamics for the electronic temperature. If the electronic temperature is lower than the steady-state temperature $T_{el}(V_T)$ (shown in Fig.~\ref{fig:fig1}(c), red curve), the temperature continues to rise. Conversely, if the electronic temperature exceeds this value, the electronic temperature of the nanowire decreases while it is superconducting. However, even in the latter case, this cooling mechanism does not immediately balance the heating stemming from the Joule effect while in the normal state. Consequently, the final temperature reached exceeds the initial switching temperature $T_S$.

As the simulations described in Eq.~(\ref{eq:Dynamics_N}) and Eq.~(\ref{eq:Dynamics_S}) are cyclically repeated, the final temperature reached at the end of each cycle increases until the cooling mechanisms balance the heating ones (see Fig.~\ref{fig:fig2}, blue to red color gradient in the upper-left scheme). This temperature is defined as the equilibrium temperature $T_{eq}$ of the device (see Fig.~\ref{fig:fig2} upper-left scheme). At this temperature, $T_{el}$ and $V_{out}$ periodically oscillate at a defined frequency $\nu_0$, which depends on the equilibrium temperature, through the switching current dependence on the temperature and the period dependence on it, shown in Eq.~(\ref{eq:Period}), and ultimately on the injection voltage $V_T$. The oscillation frequency can thus be modulated by applying a DC voltage to the injection tunnel junction. It is noteworthy that under these operating conditions, the bias current consistently exceeds the switching current at the equilibrium temperature, as $T_{eq} > T_S$, implying $I_S(T_{eq}) < I_S(T_S) = I_{bias}$. This oscillation mechanism bears similarities to those reported by \cite{toomey_frequency_2018,toomey_microwave_2017}. However, our approach offers the advantage of simple frequency modulation by applying a small voltage to the injection junction.

\section{Analysis of Performance}\label{sec:results}

\subsection{Harmonic content and frequency modulation}

\begin{figure}[h!]
\includegraphics[width=\linewidth]{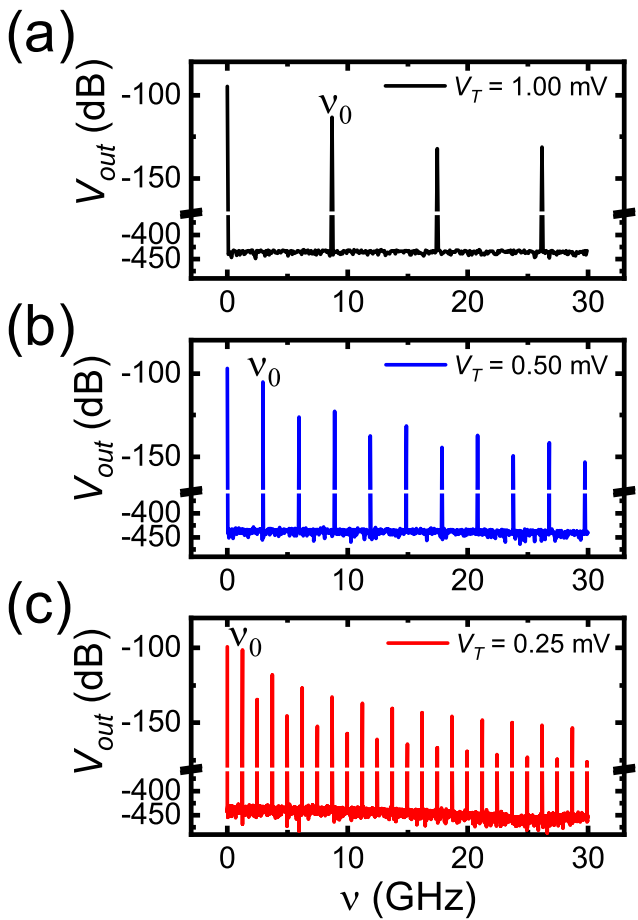}
\caption{\label{fig:fig3} Frequency spectrum of voltage oscillation vs. $V_T$ between 0  and 30 GHz, for fixed parameters $R_S = 1.00$ $\Omega$, $L_S = 0.25$ nH, and bath temperature $T_B = 0.02$ K. The spectrum is shown for three different tunnel junction voltages, showing the dependence of oscillation frequency on $V_T$.
(a) Spectrum of output voltage $V_{out}$ for $V_T = 1.00$ mV. The fundamental frequency, the inverse of the oscillation period, is $\nu_0 = 8.7$ GHz. This higher voltage results in the fastest oscillations among the three cases.
(b) Spectrum of $V_{out}$ for $V_T = 0.50$ mV. The fundamental frequency is $\nu_0 = 3.0$ GHz, showing a significant decrease in oscillation frequency compared to (a) as $V_T$ is reduced. This is the Fourier transform of the $V_{out}$ signal shown in Fig.~\ref{fig:fig1}(d).
(c) Spectrum of $V_{out}$ for $V_T = 0.25$ mV.  The fundamental frequency is $\nu_0 = 1.2$ GHz, the lowest oscillation frequency of the three cases due to the lowest applied voltage. 
As $V_T$ increases, the ratio between the first harmonic and the DC component amplitudes decreases. Higher $V_T$ raises the electronic temperature of the nanowire, reducing $I_S$ relative to $I_{bias}$. Thus, more current flows through the shunt resistor, increasing the voltage offset in $V_{out}$ and decreasing the voltage swing during oscillation. Excessive $V_T$ increases oscillation frequency but reduces amplitude until the device latches.}
\end{figure} 
To analyze the spectral characteristics of the QUISTRON, we performed a frequency analysis of the voltage oscillations for various injection junction voltages ($V_T$). Figure~\ref{fig:fig3} presents the frequency spectrum of voltage oscillations between 0 and 30 GHz, with fixed $R_S = 1.00$ $\Omega$, $L_S = 0.25$ nH, and bath temperature $T_B = 0.02$ K.
\begin{figure*}[t!]
\includegraphics[width=\linewidth]{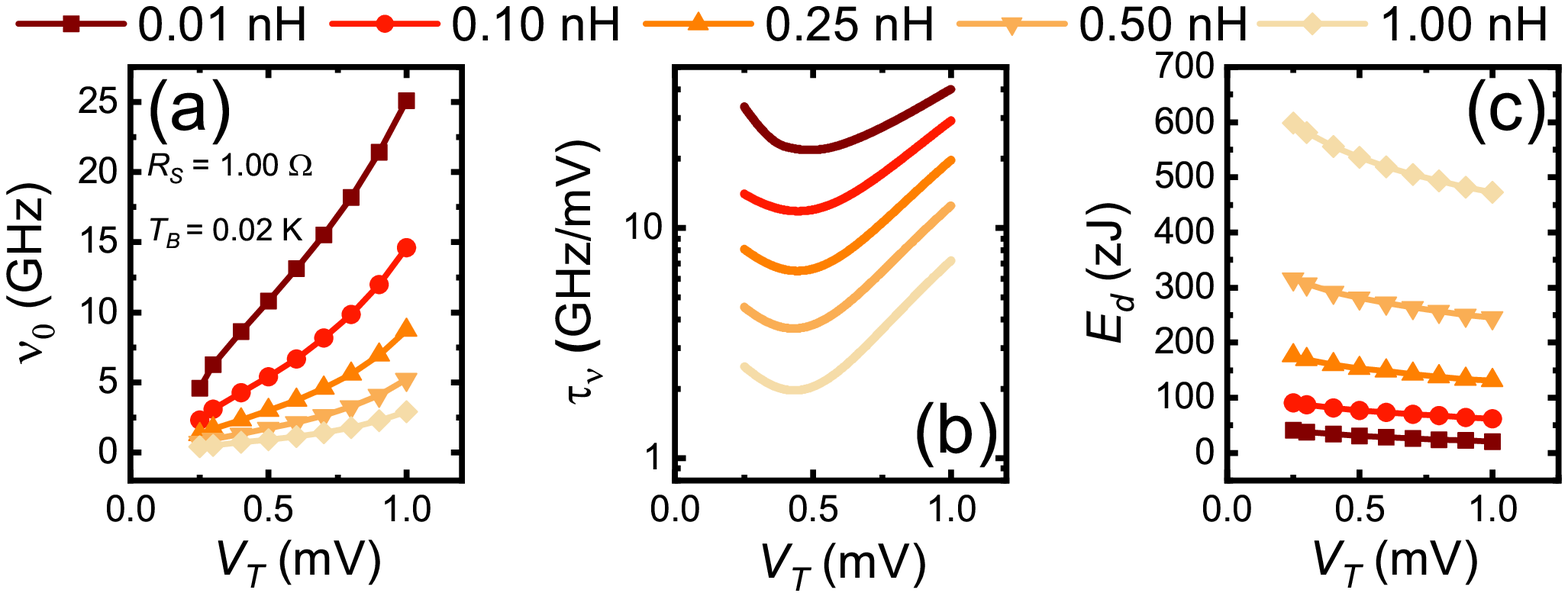}
\caption{\label{fig:fig4} Figures of merit of the QUISTRON vs. $V_T$ for different shunt inductors $L_S$, with fixed $R_S = 1.00$ $\Omega$ and $T_B = 0.02$ K.
(a) Inverse of oscillation period $\nu_0$ vs. $V_T$. $\nu_0$ increases monotonically with $V_T$. Larger $L_S$ decreases $\nu_0$ due to increased characteristic time of the RL circuit formed by nanowire and shunt.
(b) Frequency to voltage transfer function $\tau_{\nu}$, obtained by deriving the cubic spline of data in (a) for $V_T$.
(c) Total energy dissipated per period $E_d$, calculated by integrating over the oscillation period $T = 1/\nu_0$ the sum of absolute values of all heat currents, Joule power, and energy in every inductor and capacitor.}
\end{figure*}
Figure~\ref{fig:fig3} illustrates the spectrum of the output voltage $V_{out}$ for three different tunnel junction voltages, demonstrating the device voltage-controlled frequency tuning capability. In Fig.~\ref{fig:fig3}(a), with $V_T = 1.00$ mV, we observe the highest fundamental frequency of $\nu_0 = 8.7$ GHz among the three cases. This corresponds to the fastest oscillations due to the higher applied voltage. As we reduce $V_T$ to 0.50 mV in Fig.~\ref{fig:fig3}(b), the fundamental frequency decreases  to $\nu_0 = 3.0$ GHz. This substantial change in oscillation frequency shows the sensitivity of our device to the applied tunnel junction voltage. Further reducing $V_T$ to 0.25 mV, as shown in Fig.~\ref{fig:fig3}(c), results in the lowest fundamental frequency among the three cases: $\nu_0 = 1.2$ GHz. This trend clearly illustrates the inverse relationship between the applied voltage and the oscillation period. The spectra reveal not only the fundamental frequencies but also higher harmonics. This harmonic content directly results from the strongly non-sinusoidal oscillation of $V_{out}$, a typical characteristic of relaxation oscillators. 

In addition, we can notice that a DC component is present in each spectrum, as usually observed in this kind of device \cite{toomey_frequency_2018,toomey_microwave_2017}. This is because, during the oscillation, there is always a current flowing through the shunt resistor that is at least equal to $I_{bias} - I_S$, which generates a voltage $V_{offset}\simeq R_S (I_{bias} - I_S)$. It is possible to remove the offset in $V_{out}$ \cite{toomey_design_2019,toomey_superconducting_2020} by using two shunted nanowires in parallel biased in reverse one respect to the other. 
It is important to note that as the applied voltage increases, the electronic temperature of the nanowire also rises. This has a significant impact on the voltage oscillations. Specifically, it reduces the voltage swing of the oscillations. The increase in temperature leads to a decrease in the switching current relative to the bias current. Consequently, a greater fraction of the current circulates through the shunt resistor, resulting in an increased voltage offset in $V_{out}$. As the voltage offset increases while the maximum of $V_{out}$ remains constant at $R_{\parallel}I_{bias}$ (where $R_{\parallel}$ is the parallel resistance of the shunt and nanowire), the overall swing of voltage during the oscillation decreases. This behavior is illustrated in Fig.~\ref{fig:fig3}, where the ratio between the first harmonic and the zero frequency (DC component) amplitude decreases as a function of $V_T$. If the injection voltage applied is too large, the oscillation frequency rises, but the oscillation amplitude decreases until the device effectively latches.

Our research demonstrates that the QUISTRON can be adjusted across a broad frequency range, from approximately 1.2 GHz to 8.7 GHz, by modifying the tunnel junction voltage $V_T$. This adjustable feature, along with the device small size and simple control mechanism, positions it as a highly promising option for various uses in superconducting electronics and quantum information processing. The precise control of the oscillation frequency through a DC voltage provides significant advantages for circuit design and integration. Moreover, the output signal rich harmonic content could be beneficial for applications requiring non-sinusoidal waveforms or frequency multiplication.

\subsection{Impact of shunt resistor and inductor}

\begin{figure*}[t!] 
\includegraphics[width=\linewidth]{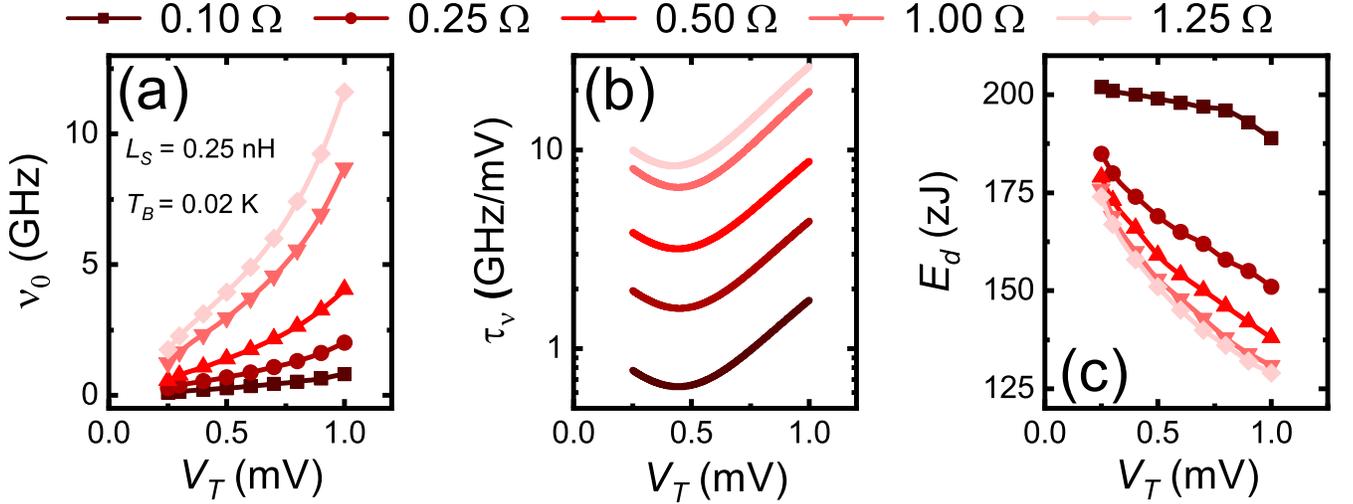}
\caption{\label{fig:fig5} Figures of merit of the QUISTRON vs. $V_T$ for different shunt resistors $R_S$, with fixed $L_S = 0.25$ nH and $T_B = 0.02$ K.
(a) Inverse of oscillation period $\nu_0$ vs. $V_T$. $\nu_0$ increases monotonically with $V_T$. Larger $R_S$ increases $\nu_0$ due to the shorter characteristic time of the RL circuit formed by nanowire and shunt.
(b) Frequency to voltage transfer function $\tau_{\nu}$, obtained by deriving the cubic spline of data in (a) for $V_T$.
(c) Total energy dissipated per period $E_d$, calculated by integrating over the oscillation period $T = 1/\nu_0$ the sum of absolute values of all heat currents, Joule power, and energy in every inductor and capacitor.}
\end{figure*}

To understand the influence of circuit parameters on the QUISTRON performance, we investigated the device behavior for different shunt inductance ($L_S$) and resistance ($R_S$) values. Figure~\ref{fig:fig4} and ~\ref{fig:fig5} present key figures of merit as functions of the tunnel junction voltage $V_T$, allowing us to analyze the impact of $L_S$ and $R_S$ separately.

Figure~\ref{fig:fig4} shows the device characteristics for selected $L_S$ values, with fixed $R_S = 1.00$ $\Omega$ and bath temperature $T_B = 0.02$ K. In Fig.~\ref{fig:fig4}(a), we observe that the inverse of the oscillation period, $\nu_0$, increases monotonically with $V_T$ for all $L_S$ values, demonstrating the voltage-controlled nature of the QUISTRON.
However, larger $L_S$ values result in lower $\nu_0$ for a given $V_T$. This behavior can be attributed to the increased characteristic time of the RL circuit formed by the nanowire and the shunt, which slows down the current redistribution process during oscillations. 
The sensitivity of the oscillator frequency to voltage changes is quantified by the frequency-to-voltage transfer function $\tau_{\nu}$, shown in Fig.~\ref{fig:fig4}(b). We obtained $\tau_{\nu}$ by calculating the derivative of a cubic spline fit to the data in Fig.~\ref{fig:fig4}(a) with respect to $V_T$. This figure of merit underlines the device oscillation frequency sensitivity to the applied voltage $V_T$.
For every mV applied on the injection junction, the oscillation frequency changes on the 1-10 GHz order. The device oscillation frequency becomes more sensitive to $V_T$ as the $L_S$ decreases because a larger $L_S$ increases the characteristic time of the RL circuit.

Figure~\ref{fig:fig4}(c) presents the total energy dissipated per period, $E_d$, as a function of $V_T$. 
We calculated $E_d$ by integrating the sum of all heat currents flowing into the nanowire, Joule power, and energy in every inductor and capacitor over one oscillation period $T = 1/\nu_0$. That can be divided into the integral of the power dissipated when the nanowire is in the normal state and the one dissipated while in the superconducting state (see Appendix \ref{sec:appendix3} for details in its estimation). We measured $E_d$ values of around 100 zJ, remarkably low and similar to those obtained in other superconducting electronic devices like RFSQ \cite{likharev_rsfq_1991}, the state-of-the-art for low-dissipation electronics.

It is worth highlighting the relationship between the oscillation frequency and the total energy dissipation per cycle ($E_d$). As the oscillation frequency increases, we notice that $E_d$ monotonically decreases. This result can be explained by considering the power consumption and the oscillation period. The device power consumption remains almost constant as we change $L_S$. A slight decrease is due to lower inductance energy, but this contribution is overall negligible. As the frequency increases, the period of each oscillation cycle decreases. $E_d$ is calculated by integrating the power consumed over one oscillation period. With a faster oscillation (shorter period) and nearly constant power consumption, we are integrating over a shorter time interval. As a result, a higher frequency leads to lower energy dissipation per cycle, even though the average power consumption remains relatively constant.
As shown in Fig.~\ref{fig:fig4}, the shunt inductance dramatically affects the device performance. By varying the inductance from 1 nH to 0.01 nH, $\nu_0$ and $\tau_{\nu}$ change by an order of magnitude. 
It is possible to adjust the oscillation frequency of the device from about 400 MHz to approximately 25 GHz by using a variable shunt inductor, such as by utilizing a SQUID or the kinetic inductance of another nanowire. This device covers all the necessary ranges for microwave applications, especially the 1-10 GHz range, which is important for quantum information applications.

Figure~\ref{fig:fig5} illustrates the device characteristics for different $R_S$ values, with fixed $L_S = 0.25$ nH and $T_B = 0.02$ K. In ~Fig.~\ref{fig:fig5}(a), we see that $\nu_0$ increases monotonically with $V_T$ for all $R_S$ values. However, in contrast to the effect of $L_S$, larger $R_S$ values lead to higher $\nu_0$ for a given $V_T$. This is due to the shorter characteristic time of the RL circuit when $R_S$ is increased, resulting in faster current redistribution and, consequently, higher oscillation frequencies. 

The frequency-to-voltage transfer function $\tau_{\nu}$ for varying $R_S$ is shown in Fig.~\ref{fig:fig5}(b), calculated using the same method as in Fig.~\ref{fig:fig4}(b). The differences in $\tau_{\nu}$ across different $R_S$ values indicate how the shunt resistance affects the device frequency tunability. Fig.~\ref{fig:fig5}(c) shows the total energy dissipated per period, $E_d$, for different $R_S$ values. As already shown in Fig.~\ref{fig:fig4}(c), in this case, we also obtained values of $E_d$ on the order of 100 zJ, with a monotonically decreasing $E_d$ as the oscillation frequency increases. However, as $R_S$ increases, so does the Joule heating, resulting in greater dissipated power. Indeed, another effect of the value of the shunt resistor $R_S$ is to change the parallel resistance of the circuit, which in turn affects the amplitude of the $V_{\text{out}}$ spikes and the Joule heating of the nanowire. For larger $R_S$, both these quantities increase. If $R_S$ becomes too large, the Joule heating in the nanowire cannot be efficiently dispersed through the cooling fin, and the nanowire electronic temperature would remain higher than the switching temperature $T_S$ even after turning off $V_T$. This occurs for shunt resistors larger than 1.25 $\Omega$. In such cases, once the nanowire is heated above $T_S$ by the quasiparticle injection, it oscillates even after $V_T$ is turned to 0 V.

The results presented in Fig.~\ref{fig:fig4} and Fig.~\ref{fig:fig5} demonstrate that both $L_S$ and $R_S$ play crucial roles in determining the performance characteristics of the QUISTRON. The shunt inductance primarily affects the oscillation frequency by modulating the characteristic time of the RL circuit. In contrast, the shunt resistance influences the frequency, the amplitude of $V_{out}$, and the power consumption. These findings provide valuable insights for optimizing the device performance for specific applications. For instance, if a higher oscillation frequency is desired, one could opt for a lower $L_S$ or a higher $R_S$. Furthermore, the voltage-dependent behavior of $\nu_0$, $\tau_{\nu}$, and $E_d$ across different $L_S$ and $R_S$ values suggests that the device operating point can be finely tuned to meet specific requirements in terms of frequency, sensitivity, and energy efficiency. This flexibility makes the QUISTRON a versatile component for various superconducting electronic applications.

\begin{figure}[h!]
\includegraphics[width=\columnwidth]{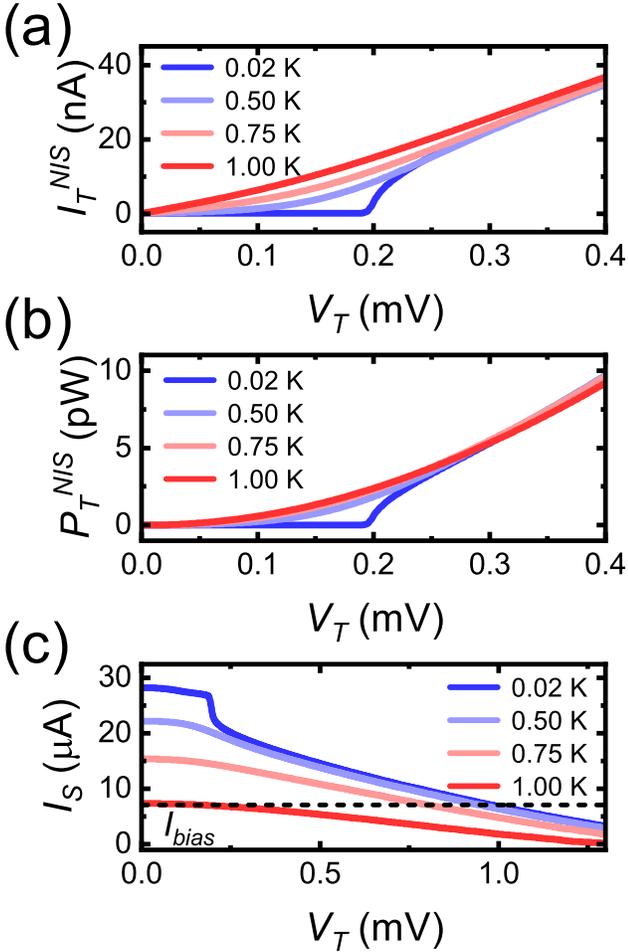}
\caption{\label{fig:fig6} Impact of bath temperature $T_B$ (ranging from 0.02 K to 1.00 K) on a NIS (normal metal-insulator-superconductor) tunnel junction characteristics and the associated nanowire switching current ($I_S$). (a) Charge current ($I_T^{NIS}$) vs. applied voltage ($V_T$) curves. At low $T_B$, the subgap current is negligible, showing electrical isolation. Current increases rapidly at $V_T \simeq \Delta_1/e = 0.2$ mV, where $\Delta_1$ is the superconducting gap energy. Higher $T_B$ smooths the curves and increases the subgap current due to thermal quasiparticle excitation. (b) Heat current through the NIS junction $P_T^{NIS}$ vs. $V_T$ curves, showing thermal isolation at low $T_B$ for subgap voltages and increased heat flow and smoother curves at higher $T_B$. (c) Nanowire switching current ($I_S$) vs. $V_T$ curves, critical for device operation. At $T_B = 0.02$ K, $I_S$ drops sharply at $V_T \simeq \Delta_1/e = 0.2$ mV due to the sharp increase in heat current near $\Delta_1/e$. Higher $T_B$ smooths these curves, mirroring the smoothing of the heat current vs. $V_T$ curves in (b). Dashed black line indicates $I_{bias}$ used in simulations shown in Fig.~\ref{fig:fig7}.}
\end{figure}

\subsection{Impact of bath temperature} 

The performance of the QUISTRON is significantly influenced by the operating temperature. To investigate this dependence, we analyzed the device characteristics across a range of bath temperatures ($T_B$) from 0.02 K to 1.00 K. Figure~\ref{fig:fig6} illustrate the effects of temperature on the NIS (Normal metal-Insulator-Superconductor) tunnel junction characteristics and the nanowire switching current. 
\begin{figure*}[t!]
\includegraphics[width=\linewidth]{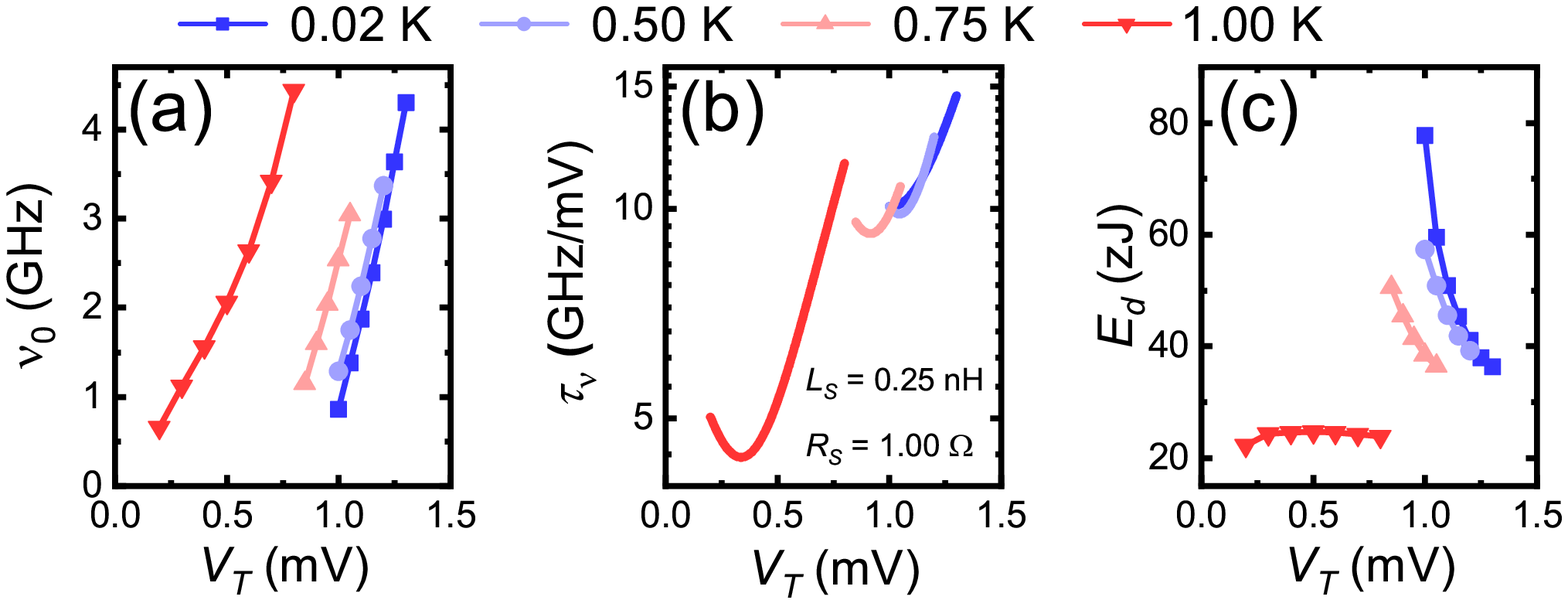}
\caption{\label{fig:fig7} Figures of merit of the QUISTRON vs. $V_T$ for different bath temperatures $T_B$, with fixed $L_S = 0.25$ nH and $R_S = 1.00$ $\Omega$.
(a) Inverse of oscillation period $\nu_0$ vs. $V_T$. $\nu_0$ increases monotonically with $V_T$. Higher $T_B$ shifts nanowire operation voltage towards lower values, as less voltage (and power) is needed to reduce switching current below bias current at higher $T_B$.
(b) Frequency to voltage transfer function $\tau_{\nu}$, obtained by deriving the cubic spline of data in (a) for $V_T$.
(c) Total energy dissipated per period $E_d$, calculated by integrating over the oscillation period $T = 1/\nu_0$ the sum of absolute values of all heat currents, Joule power, and energy in every inductor and capacitor.}
\end{figure*}

In Fig.~\ref{fig:fig6}(a), we observe the charge current ($I_T$) versus applied voltage ($V_T$) characteristics. At low $T_B$, the subgap current is negligible, indicating effective electrical isolation. A sharp increase in current occurs at $V_T \simeq \Delta_1/e = 0.2$ mV, where $\Delta_1$ is the superconducting gap energy. As $T_B$ increases, the I-V curves become smoother, and the subgap current rises due to the thermal excitation of quasiparticles. The heat current $P_T^{NIS}$ versus $V_T$ curves in Fig.~\ref{fig:fig6}(b) show similar temperature-dependent behavior. At low $T_B$, there is thermal isolation for subgap voltages. Higher bath temperatures lead to increased heat flow and smoother curves. Figure~\ref{fig:fig6}(c) illustrates the nanowire switching current ($I_S$) versus $V_T$, which is critical for device operation. At $T_B = 0.02$ K, $I_S$ drops sharply at $V_T \simeq \Delta_1/e = 0.2$ mV, corresponding to the sharp increase in heat current near $\Delta_1/e$. As $T_B$ increases, these curves become smoother, mirroring the behavior of the heat current curves in Fig.~\ref{fig:fig6}(b). The dashed black line indicates the bias current $I_{bias}$ used in simulations. It was chosen to be smaller than the switching current at 1 K to use the same bias current in each simulation. It should be noted that this choice of bias current is not optimal for every bath temperature because a lower bias current than the switching current means that a larger injected power is needed to activate the oscillation, thus decreasing the efficiency of the device.

The results presented in Fig.~\ref{fig:fig6} and Fig.~\ref{fig:fig7} show the impact of bath temperature on the device. The temperature-dependent behavior of the NIS junction, particularly the smoothing of $I-V$ characteristics and the shift in nanowire switching current, directly influences the device operating range and performance. The device can operate at lower voltages and bias currents at higher temperatures, which could be advantageous for low-power applications. However, this comes at the cost of increased subgap currents and heat flow, potentially affecting the device energy efficiency. The temperature dependence of $\nu_0$, $\tau_{\nu}$, and $E_d$ suggests that the device is robust and predictable under bath temperature variations, as its functioning was simulated up to a bath temperature of $T_B = 1$ K. In addition, this temperature sensitivity could be exploited in specific applications, such as bolometry or temperature sensing.

Figure~\ref{fig:fig7} presents the device figures of merit as functions of $V_T$ for different bath temperatures, with fixed $L_S = 0.25$ nH and $R_S = 1.00$ $\Omega$. Fig.~\ref{fig:fig7}(a) shows that the inverse of the oscillation period ($\nu_0$) increases monotonically with $V_T$ for all $T_B$. However, higher bath temperatures shift the nanowire operation voltage towards lower values. This shift occurs because less voltage (and power) is needed to reduce the switching current below the bias current at higher $T_B$, as evidenced by the $I_S$ vs. $V_T$ curves in Fig.~\ref{fig:fig6}(c). The frequency-to-voltage transfer function ($\tau_{\nu}$), shown in Fig.~\ref{fig:fig7}(b), is obtained by calculating the derivative of a cubic spline fit to the data in Fig.~\ref{fig:fig7}(a) for $V_T$. The temperature dependence of $\tau_{\nu}$ shows how the device frequency tunability changes with bath temperature. As already seen in Fig.~\ref{fig:fig7}(a), the bath temperature main effect is shifting the nanowire operation voltage towards lower values.

Notably, the largest swing in $\tau_{\nu}$ is obtained at $T_B = 1$ K. The reason is that the bias current chosen to perform the simulation is slightly lower than the switching current at 1 K, as shown in Fig.~\ref{fig:fig6}(c). This is the optimal configuration for the device because it allows for a low switching voltage and large operation voltages. Figure~\ref{fig:fig7}(c) presents the total energy dissipated per period ($E_d$) as a function of $V_T$ for different $T_B$. We calculated $E_d$ by integrating the sum of absolute values of all heat currents, Joule power, and energy in every inductor and capacitor over one oscillation period $T = 1/\nu_0$.

As the bath temperature $T_B$ increases, we observe a decrease in the total energy dissipated per period, $E_d$. This inverse relationship can be attributed to the reduced power required from the quasiparticle-injection tunnel junction to switch the nanowire. The reduction in power requirement stems from the decrease in switching current with increasing $T_B$. However, this analysis does not fully capture an essential aspect of the device behavior: the decrease in voltage swing as $T_B$ increases, a phenomenon previously explained and illustrated in Fig.~\ref{fig:fig3}. While the device remains operational at higher temperatures, this temperature dependence has significant implications for its performance. Specifically, the AC power output of the device decreases with increasing $T_B$.

\section{Conclusions} \label{sec:conclusions}

In conclusion, our investigations demonstrate the potential of an original relaxation oscillator design: the QUISTRON. This device comprises a superconducting nanowire shunted by a resistor and an inductor, with its operation controlled by a voltage applied to a tunnel junction positioned on the nanowire. The oscillation frequency is modulated via a DC voltage applied to a normal metal-insulator-superconductor (NIS) tunnel junction, inducing quasiparticle injection and localized heating of the nanowire. This mechanism effectively modulates the nanowire switching current, initiating oscillations when it falls below a DC bias current.

Numerical simulations have confirmed the device functionality and explored its performance characteristics. The results are promising, revealing an oscillation frequency range of 1.2 GHz to 8.7 GHz—ideal for microwave applications for a specific configuration of shunt resistor and inductor ($R_S = 1$ $\Omega$, $L_S = 0.25$ nH). Moreover, the QUISTRON exhibits ultra-low energy dissipation of $\sim100$ zJ per cycle and maintains operational stability across a broad temperature range (20 mK to 1 K). These attributes underscore the QUISTRON potential as a compact, tunable, and localized superconducting oscillator. Its straightforward DC voltage control mechanism and compatibility with circuit integration render it a compelling candidate for diverse applications in quantum information processing, microwave technology, and ultra-low-power electronics.

Notably, any mechanism to decrease the switching current in a superconducting channel shunted by a resistor could be exploited to design a superconducting relaxation oscillator. Alternative architectures for voltage-controlled superconducting relaxation oscillators could be developed using different principles. These include superconductor-normal metal-superconductor (SNS) junctions with semiconducting channels controlled by gating effects \cite{PhysRevB.93.155402, PhysRevResearch.3.L022005} and gate-controlled superconductivity in metallic superconductors \cite{de_simoni_metallic_2018, paolucci_field-effect_2019}. Such diverse approaches highlight the rich potential for innovation in 
superconducting relaxation oscillators.
Future research endeavors will focus on experimentally validating the proposed design and further optimizing its performance characteristics.

\section*{Acknowledgements} \label{sec:acknowledgements}

We acknowledge the EU’s Horizon 2020 Research
and Innovation Framework Programme under Grants No.
964398 (SUPERGATE), No. 101057977 (SPECTRUM),
and the PNRR MUR project PE0000023-NQSTI for partial
financial support.

\appendix
\section{Estimation of heat currents} \label{sec:appendix1}

The injection junction charge and heat currents flowing into the nanowire when it is superconducting, shown in Fig.~\ref{fig:fig1}(b), are expressed by \cite{giazotto_opportunities_2006}:
\begin{align}
    I_T^{NIS} &= \frac{1}{e R_T}\int_{-\infty}^{\infty}d\epsilon\,\mathcal{N}_{S_1} (\epsilon, T_{el}) \nonumber \\ 
    &\quad \times [f_0(\epsilon - e V_T, T_B) - f_0(\epsilon, T_{el})], \\ \nonumber \\
    P_T^{NIS} &= \frac{1}{e^2 R_T}\int_{-\infty}^{\infty}d\epsilon\,\epsilon\,\mathcal{N}_{S_1}(\epsilon, T_{el}) \nonumber \\ 
    &\quad \times [f_0(\epsilon - e V_T, T_B) - f_0(\epsilon, T_{el})],
\end{align}
where $e$ is the elementary charge, $R_T$ is the tunnel junction resistance, $V_T$ is the voltage applied to the junction, $T_{el}$ is the electronic temperature of the nanowire, $T_B$ is the bath temperature, $f_0(E, T)=1/[\exp{(E/k_B T)}+1]$ is the Fermi-Dirac distribution of the quasiparticles at energy $E$ and electronic temperature $T$, and $\mathcal{N}_{S_1}(E, T) = \left|\Re{\left[E+i\Gamma_1/\sqrt{(E+i\Gamma_1)^2 - \Delta_1^2(T)}\right]}\right|$ is the (smeared by the Dynes parameter $\Gamma_1=10^{-3}\Delta_1$) BCS density of states (DOS) of the $S_1$ superconductor normalized at the Fermi level \cite{pekola2004limitations}. The charge and heat current when the nanowire is in the normal state, respectively $I_T^{NIN}$ and $P_T^{NIN}$, are the same expressions in which the normalized DOS is set to be normal metal one, which is constantly equal to 1.
The cooling fin junction heat current flowing out of the nanowire when it is in the superconducting is expressed by:
\begin{align}
    P_C^{NIS} &= -\frac{1}{e^2 R_C} \int_{-\infty}^{\infty} d\epsilon\, \epsilon\, \mathcal{N}_{S_1}(\epsilon, T_{el}) \nonumber \\
    &\quad \times [f_0(\epsilon, T_B) - f_0(\epsilon, T_{el})],
\end{align}
where $R_C$ is the tunnel junction resistance. The heat current when the nanowire is in the normal state, $P_C^{NIN}$, is given by the same expression with the normalized DOS set to that of a normal metal, which is constantly equal to 1. The absence of a superconducting gap, which exponentially suppresses the subgap DOS, amplifies the heat current flowing in the cooling fin while the nanowire is in the normal state, introducing a feedback control on the electronic temperature of the nanowire that depends on its state. 

The lateral banks junction heat current flowing out of the nanowire when it is superconducting is expressed by \cite{giazotto_opportunities_2006}:
\begin{align}
    2P_J^{SIS} &= -\frac{2}{e^2 R_J} \int_{-\infty}^{\infty} d\epsilon\, \epsilon\, \mathcal{N}_{S_1}(\epsilon, T_{el}) \mathcal{N}_{S_2}(\epsilon, T_B) \nonumber \\
    &\quad \times [f_0(\epsilon, T_B) - f_0(\epsilon, T_{el})],
\end{align}
where $R_J$ is the resistance of each tunnel junction, and $\mathcal{N}_{S_2}(E, T) = \left|\Re{\left[(E+i\Gamma_2)/\sqrt{(E+i\Gamma_2)^2 - \Delta_2^2(T)}\right]}\right|$ is the (smeared by the Dynes parameter $\Gamma_2=10^{-3}\Delta_2$) BCS density of states (DOS) of the $S_2$ superconductor normalized at the Fermi level \cite{pekola2004limitations}.

When the nanowire is in the normal state, the series of the two junctions and the nanowire effectively acts as a biased SINIS structure. In this case, the heat current also depends on the voltage applied across the structure, particularly on the voltage drop across the two tunnel junctions \cite{giazotto_opportunities_2006, giazotto_josephson_2005, savin_cold_2004}:
\begin{align}
    P_J^{SINIS} &= 2P_J^{NIS} = -\frac{2}{e^2 R_J} \int_{-\infty}^{\infty} d\epsilon\, \epsilon\, \mathcal{N}_{S_2}(\Tilde{\epsilon}, T_B) \nonumber \\
    &\quad \times [f_0(\Tilde{\epsilon}, T_B) - f_0(\epsilon, T_{el})],
\end{align}
where $\Tilde{\epsilon} = \epsilon - V_{out}(R_J/R_N)$, and $V_{out}(R_J/R_N)$ is the voltage drop across each tunnel junction, which depends on the ratio between the junction resistance $R_J$ and the total series resistance of the two junctions and the nanowire in the normal state $R_N=R_{NW}+2R_J$.

The heat current flowing out of the nanowire due to the electron-phonon coupling when the nanowire is in the normal state is expressed by \cite{giazotto_opportunities_2006}:
\begin{align}
    P_{e-ph}^{N} &= \Sigma V \left(T^5_{el} - T^5_B\right),
\end{align}
where $\Sigma$ is the electron-phonon coupling constant, which for aluminum at sub-Kelvin temperatures is $\Sigma=0.3\times 10^9$ W/(m$^3$K$^5$) \cite{giazotto_opportunities_2006,meschke_electron_2004}, and $V$ is the nanowire volume.

The heat current flowing out of the nanowire due to the electron-phonon coupling when the nanowire is superconducting is expressed by \cite{timofeev2009recombination,PhysRevLett.111.147001}:
\begin{align}
    P_{e-ph}^{S} &= \frac{\Sigma V}{24 \zeta(5) k_B^5} \int_{0}^{\infty}d\epsilon\,\epsilon^3 [n(\epsilon, T_{el}) - n(\epsilon, T_{B})] \nonumber \\
    &\quad \times \int_{-\infty}^{\infty}dE\, \mathcal{N}_{S_1}(E, T_{el}) \mathcal{N}_{S_1}(E + \epsilon, T_{el}) \nonumber \\
    &\quad \times \left[1 - \frac{\Delta_1^2(T_{el})}{E(E + \epsilon)}\right] [f_0(E, T_{el}) - f_0(E + \epsilon, T_{el})],
\end{align}
where $\zeta$ is the Riemann zeta function, $n(\epsilon, T) = 1/[\exp{(\epsilon/k_B T)} - 1]$ is the Bose-Einstein distribution of the nanowire phonons at energy $\epsilon$ and temperature $T$, and $\Delta_1(T)$ is the nanowire superconducting gap. The superconducting gap suppresses the electron-phonon coupling at low temperatures ($T \ll \Delta_1 (T)/k_B$), thus $P_{e-ph}^{S} < P_{e-ph}^{N}$.

\section{Estimation of nanowire heat capacity and electron-phonon length} \label{sec:appendix2}

The nanowire electronic heat capacity while in the normal state $C_N$ was estimated as:
\begin{align}
    C_N(T)=\frac{\pi^2}{3}k_B^2 T V D(E_F),
\end{align}
where $k_B$ is the Boltzmann constant, $V$ is the nanowire volume, and $D(E_F)=1.45\times 10^{47}$ 1/(Jm$^3$) is the aluminum density of state at the Fermi energy $E_F=11.6$ eV.

The nanowire electronic heat capacity while in the superconducting state $C_S$ was calculated starting from a fermionic system entropy relation:
\begin{equation}
    S(T) = -2k_B\sum_k [f_k \ln(f_k)+ (1-f_k)\ln(1-f_k)],
\end{equation}
where $f_k$ is the Fermi distribution at energy $E=\sqrt{\epsilon_k^2+\Delta_1^2(T)}$ and temperature $T$. Then, the 
heat capacity of a BCS superconductor is \cite{tinkham2004introduction}:
\begin{align}
    C_S(T) &= T \frac{\partial S(T)}{\partial T} =
    T \sum_k \frac{\partial S}{\partial f_k}\frac{\partial f_k}{\partial T}\nonumber\\
    & = \frac{2}{k_B T^2}\sum_k \frac{e^{\frac{\sqrt{\epsilon_k^2+\Delta_1^2(T)}}{k_B T}}}{\bigg[e^{\frac{\sqrt{\epsilon_k^2+\Delta_1^2(T)}}{k_B T}}+1\bigg]^2}\nonumber\\
    &\quad\times\bigg[\epsilon_k^2 + \Delta_1^2(T) - \frac{T}{2}\frac{d}{dT}\Delta_1^2(T)\bigg],
\end{align}
which in the continuous limit can be written as:
\begin{align}
   C_S(T)&=\frac{V D(E_F)}{k_B T^2}\int_{-\infty}^{\infty}\frac{e^{\frac{\sqrt{\epsilon^2+\Delta_1^2(T)}}{k_B T}}}{\bigg[e^{\frac{\sqrt{\epsilon^2+\Delta_1^2(T)}}{k_B T}}+1\bigg]^2}\nonumber \\
   &\quad \times \bigg[\epsilon^2+\Delta_1^2(T)-T\Delta_1(T)\frac{d\Delta_1(T)}{dT}\bigg]d\epsilon.
\end{align}
The nanowire electron-phonon scattering length while in the normal state, $l_{e\text{-}ph}^N$, was estimated as:
\begin{align}
    l_{e\text{-}ph}^N = \sqrt{D_{Al} \tau_{e\text{-}ph}^N},
\end{align}
where $D_{Al} = 1/[e^2 \rho_{Al} D(E_F)]$ is the aluminum diffusion constant, $\rho_{Al}$ is the resistivity of an aluminum film 20 nm thick \cite{PhysRevB.26.3648}, and $\tau_{e\text{-}ph}^N$ is the electron-phonon scattering time when the nanowire is in the normal state, which is:
\begin{align}
    \tau_{e\text{-}ph}^N &= \frac{C_N}{G_N}, \\
    G_N &= \frac{dP_{e\text{-}ph}^N}{dT},
\end{align}
where $C_N$ is the electronic heat capacitance of the nanowire, and $G_N$ is the electron-phonon thermal conductance in the normal state.

The electron-phonon scattering length $l_{e\text{-}ph}^N$ is a monotonically decreasing function of temperature. We estimated it at the highest possible operation temperature, which is the critical temperature of the aluminum nanowire $T_{c1} = 1.3$ K. It is $l_{e\text{-}ph}^N(T_{c1}) \simeq 17\ \mu$m, which is much larger than the length of the nanowire $l = 1.2\ \mu$m. This means that the temperature dynamics is homogeneous along the entire nanowire while in the normal state.

The nanowire electron-phonon scattering length while in the superconducting state, $l_{e\text{-}ph}^S$, was estimated as:
\begin{align}
    l_{e\text{-}ph}^S = \sqrt{D_{Al} \tau_{e\text{-}ph}^S},
\end{align}
where $\tau_{e\text{-}ph}^S$ is the electron-phonon scattering time when the nanowire is superconducting, which is:
\begin{align}
    \tau_{e\text{-}ph}^S &= \frac{C_S}{G_S}, \\
    G_S &= \frac{dP_{e\text{-}ph}^S}{dT}.
\end{align}

The electron-phonon scattering length $l_{e\text{-}ph}^S$ is a monotonically decreasing function of temperature. We estimated it at the highest possible operation temperature, which is the critical temperature of the aluminum nanowire $T_{c1} = 1.3$ K. It is $l_{e\text{-}ph}^S(T_{c1}) \simeq 200\ \mu$m, which is much larger than the length of the nanowire $l = 1.2\ \mu$m. This means that the temperature dynamics is also homogeneous along the entire nanowire while in the superconducting state. These estimations justify the assumption we made to neglect the spatial gradients of the temperature along the nanowire.

\section{Estimation of nanowire retrapping current, kinetic inductance and energy dissipation} \label{sec:appendix3}

The nanowire retrapping current $I_R$ was estimated from the thermal balance between the self-heating due to the Joule power generated while the nanowire is in the normal state and the other heat currents flowing in or from the nanowire, as in \cite{skocpol_self-heating_1974, tinkham_hysteretic_2003, dane_self-heating_2022}. The Joule power dissipated in the nanowire at the retrapping current is $P_{Joule}= [R_{NW}+2(R_J/2)]I_R^2$, where we assumed that half of the power dissipated on the insulating barriers of resistance $R_J$ is dissipated in the nanowire. Thus, the retrapping current is expressed by the relation:
\begin{align}
I_R=\sqrt{\frac{P_C^{NIN}+P_J^{SINIS}+P_{e-ph}^N-P_T^{NIN}}{R_{NW}+R_J}},
\end{align}
where we assumed that the electronic temperature was equal to the critical temperature, as done in \cite{skocpol_self-heating_1974}. However, this assumption is not completely correct in our device, as we have seen that the actual electronic temperature of the nanowire is variable during the operation, and it is always smaller than $T_{c1}$.

The device kinetic inductance $L_K$ was estimated as the sum of the nanowire kinetic inductance $L_K^{NW}$ and the two Josephson junctions kinetic inductances $L_K^J$. We estimated $L_K^{NW}$ from the BCS expression for a superconducting stripe \cite{battisti_demonstration_2024}:
\begin{align}
    L_K^{NW}(T) = \frac{\hbar R_{NW}}{\pi \Delta_1(T) \tanh{\bigg[\frac{\Delta_1(T)}{k_B T}\bigg]}},
\end{align}
while $L_K^J$ was estimated from the linearized Josephson expression:
\begin{align}
    L_K^J(T) = \frac{\Phi_0}{2 \pi I_c(T)},
\end{align}
where $\Phi_0 = \frac{h}{2e} \simeq 2.07$ fWb is the superconducting flux quantum.

The total energy dissipated per cycle $E_d$ was calculated by integrating the sum of the heat currents flowing into the nanowire, Joule power, and energy in every inductor and capacitor over one oscillation period $T = 1/\nu_0$:
\begin{align}
E_d&=\bigg(\int_{0}^{t_R}[V_{out}(t)I_{bias}+P_T^{NIN}]dt \nonumber \\
&\quad+\frac{1}{t_R}\int_{0}^{t_R}\bigg[\frac{1}{2}C_J^{eff} V_{out}^2(t)+\frac{1}{2}L_S [I_{bias}-I_{NW}(t)]^2\bigg]dt\bigg)\nonumber \\
&\quad+\bigg(\int_{t_R}^{T} [V_{out}(t)I_{bias}+P_T^{NIS}]dt \nonumber \\
&\quad+\frac{1}{T-t_R}\int_{t_R}^{T}\bigg[\frac{1}{2}C_J^{eff} V_{out}^2(t)+\frac{1}{2}L_K I_{NW}^2(t) \nonumber \\
&\quad+\frac{1}{2}L_S[I_{bias}-I_{NW}(t)]^2\bigg]dt\bigg).
\end{align}
This expression can be divided into the integral of the power dissipated when the nanowire is in the normal state (first two terms) and the power dissipated while in the superconducting state (last two terms), where $C_J^{eff} = (1/C_J + 1/C_J)^{-1} = C_J/2$ is the effective capacitance given by the series of the two junctions between the niobium banks and the aluminum nanowire and $t_R$ is the time during which the nanowire stays in the normal state. This is the time necessary for the current flowing in the nanowire to transition from the switching current to the retrapping current value with a time constant $\tau_1$. The junctions capacitance is estimated using the expression $C_J = c_s A_J = 100$ fF, where $c_s = 50 \text{ fF}/\mu\text{m}^2$ is the specific capacitance and $A_J = 2 \mu\text{m}^2$ is the junctions surface area.

\end{document}